\documentclass[openacc]{rstransa}
\usepackage[numbers]{natbib}
\usepackage{graphicx}
\usepackage{tikz}
\usepackage{amsmath}
\usepackage{cite}
\usetikzlibrary{arrows.meta, decorations.pathreplacing, positioning}

\jname{rsta}
\Journal{Phil. Trans. R. Soc}
\titlehead{Research}

\begin{document}

\title{Dynamics of diffusive-convective staircases in the ocean}
\author{Mary-Louise Timmermans$^{1}$ and Jeffrey R. Carpenter$^{2}$}

\address{$^{1}$Department of Earth and Planetary Sciences, Yale University, New Haven, CT USA\\
$^{2}$Institute of Coastal Ocean Dynamics, Helmholtz-Zentrum Hereon, Geesthacht, Germany}

\subject{Fluid Mechanics, Oceanography, Physical Oceanography}

\corres{Mary-Louise Timmermans\\
\email{Mary-Louise.Timmermans@yale.edu}}

\begin{abstract}
Diffusive-convective (DC) staircases in the ocean are observed across a wide range of settings, but their formation, structure, and persistence are not fully understood. Theories for DC staircases are reviewed to identify mechanisms governing their development and evolution. Staircase evolution through layer merging and possibly interface splitting, including the relationship to background turbulence, is assessed. Oceanographic examples illustrate the variety of settings in which DC staircases are found, and how they can persist under weak turbulence but are disrupted when turbulence becomes sufficiently strong. Key open questions are identified, highlighting the challenge of linking small-scale processes to the large-scale coherence and persistence of DC staircases in the ocean.
\end{abstract}

\begin{fmtext}
\section{Introduction}

Ocean mixing plays a fundamental role in the climate system by controlling the distribution and uptake of heat and carbon, regulating nutrient availability, shaping water-mass transformations, and influencing the dissipation of energy. Double-diffusive convection is a type of ocean mixing that can take place when either temperature $T$ or salinity $S$ contributes a destabilizing effect to the overall stable stratification \cite{radko_double-diffusive_2013}. The background potential energy (BPE) associated with the sorted density field, which is established by the unstable configuration of either $T$ or $S$, provides the energy source for convection. This energy can be released because $T$ and $S$ have different rates of molecular diffusion, $\kappa_T$ and $\kappa_S$, where $\kappa_T$ is about 100 times larger than $\kappa_S$. A stable stratification characterized by cooler and fresher water below warmer and saltier water can support the salt fingering (SF) form of double-diffusive convection. A stably stratified setting with warmer and saltier water below cooler and fresher water can support the diffusive convective (DC) form. We restrict our focus here to the DC regime. 


\end{fmtext}
\maketitle

\begin{figure}[!h]
\centering
\includegraphics[width=4in]{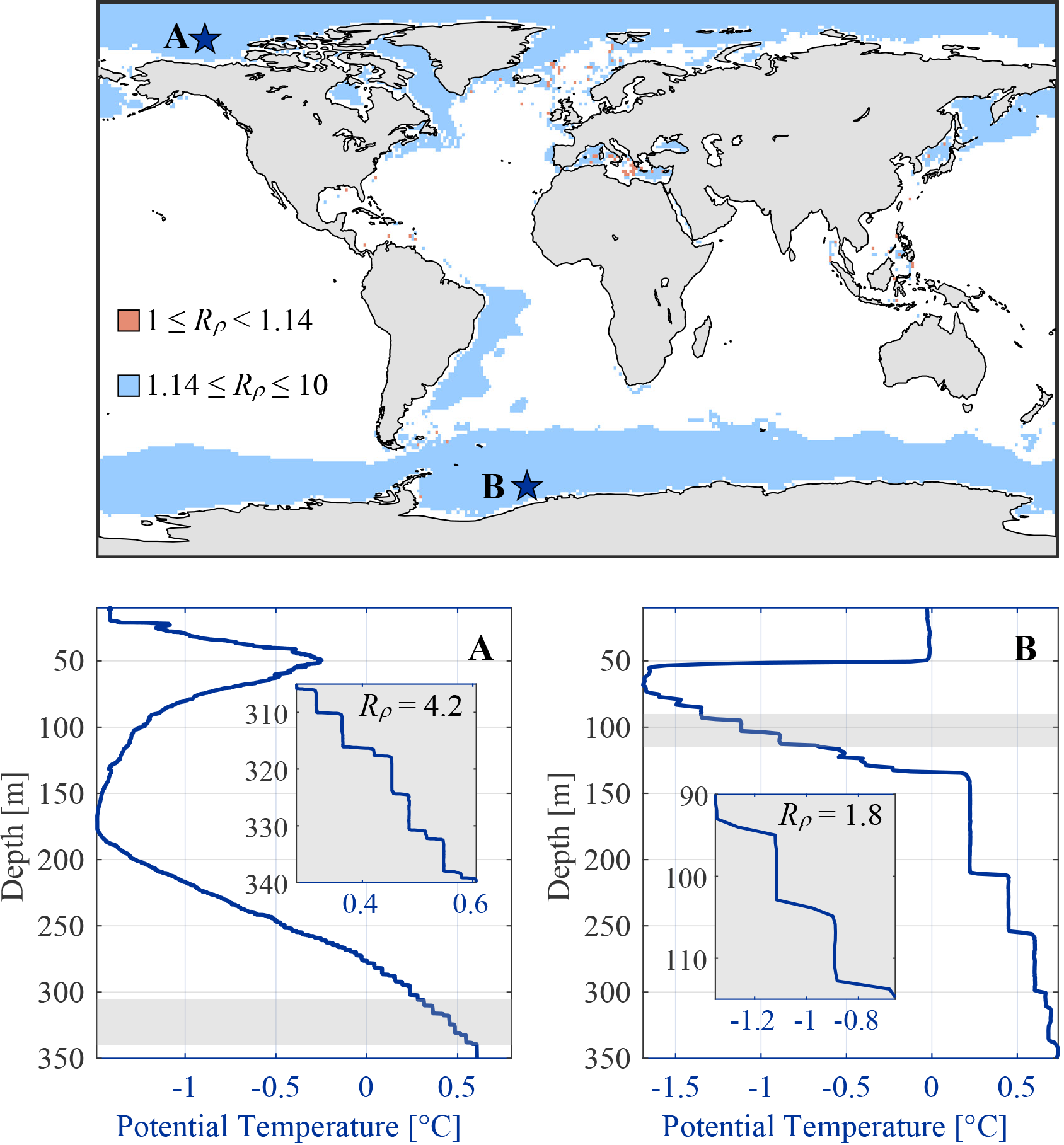}
\caption{Top: Ocean regions potentially susceptible to diffusive convection (DC), shown by density ratio $R_{\rho}$ regimes indicating where a portion of the water column (between 50~m and 2000~m depth) has $R_{\rho}$ within the ranges shown. The lower values represent the linearly unstable regime, while the higher indicate the general range where DC has been observed in both laboratory and oceanic settings \cite{kelley_diffusive_2003}. $R_{\rho}$ is computed from centered differences over the vertical grid spacing (5~m above 100~m depth, increasing to 50~m at 500~m depth and below); while this bulk value is useful for regional characterization, convective release is ultimately related to local interfacial gradients.
Bottom: Potential temperature vs. depth profiles from the upper water column of the Arctic Ocean (A) and the Southern Ocean (B), with locations shown as blue stars in the map above; the insets indicate $R_{\rho}$ based on linear gradients over the inset depth range. The map uses World Ocean Atlas 2023 Data  \cite{reagan2024woa}, objectively analyzed temperature and salinity means based on ocean profiles from the World Ocean Database (WOD; \cite{Mishonov2024WOD}); profiles are from the WOD. van der Boog et al. \cite{van_der_boog_double-diffusive_2021} provide a global synthesis of staircase prevalence in observations.}
\label{rrhomap}
\end{figure}

A key metric that in part characterizes the susceptibility of an ocean region to double-diffusive convection is the density ratio, $R_{\rho} = (\beta dS/dz)/(\alpha dT/dz)$ where $dS/dz$ and $dT/dz$ are the background vertical gradients of $S$ and $T$, $\alpha$ is the coefficient of thermal expansion and $\beta$ is the coefficient of haline contraction. In ocean regions characterized by specific density ratios, and where turbulence is not too strong, double-diffusive convection often (although not always) manifests as staircase structures characterized by layers that are well-mixed in temperature and salinity separated by high gradient interfaces (Figure~\ref{rrhomap}). Linear stability analysis of linear profiles of $T$ and $S$  \cite{veronis1965finite} indicates that the DC case in the ocean is unstable to infinitesimal disturbances for only the narrow range $R_{\rho} < 1.14$. However, DC staircases are observed in ocean regions where $R_{\rho}$ is significantly larger than this upper bound of instability (Figure \ref{rrhomap}). 

DC staircases are observed in a range of ocean settings, perhaps most commonly in the high-latitude, salinity-stratified oceans (Figure \ref{rrhomap}, profiles A and B). Alongside the basic stratification requirements (warmer, saltier water underlying cooler, fresher water), DC staircases can be associated with and influenced by, for example, geothermal heating \cite{timmermans_thermohaline_2003}, anomalous water mass inflows \cite{umlauf_diffusive_2018}, basin-scale gyre circulation \cite{shibley_beaufort_2022}, thermohaline intrusions \cite{bebieva_relationship_2017}, or ocean eddies \cite{bebieva_examination_2016}. For similar background conditions, coherent staircases may be either present and effectively permanent or absent entirely for reasons that are not fully understood \cite{shibley_spatial_2017}. Even the presence of internal waves, assumed to be disruptive, does not always correlate with changes in the persistence or structure of a staircase \cite{boury_observations_2022}. Although the dynamics of DC staircases seem to vary across the range of conditions in which they are found, the presence of staircases under certain conditions suggests the existence of some unifying principles.

This review will explore the underlying mechanisms relevant to DC staircase formation, evolution and persistence to clarify how the dominant physical mechanisms play a role in ocean observations. The numerous open questions that still exist on the topic will be highlighted. The review and citations are not exhaustive due to length constraints, but instead identify representative examples and knowledge gaps to motivate future work.

The first part of the paper establishes the theoretical foundation for DC staircases, and the latter part applies this framework to ocean observations. The next section presents a discussion of the transfer of energy in a double-diffusive system as explaining these pathways is essential for understanding how persistent staircase structures can be maintained in DC systems. In section \ref{interface}, the structure and processes in the DC interface separating staircase mixed layers are examined. Key mechanisms proposed for the initial formation of DC staircases are described in section \ref{mechanisms}. Section \ref{evolution} explores how staircases evolve after  formation, and the role of turbulence. Ocean examples are provided in section \ref{ocean}, and section \ref{summary} summarizes and presents open questions and directions for future research.

\section{Energy pathways in DC systems}
\label{energy}

In purely DC systems that have no external energy input, the energy source derives from the Background Potential Energy (BPE, the minimum potential energy state of the system) associated with the sorted density field, which is determined by the unstable temperature configuration. Middleton and Taylor \cite{middleton_general_2020} relate the release of BPE to the sign of the diapycnal buoyancy flux. This framework was extended by Tailleux \cite{tailleux_negative_2024} who considers energy pathways in a closed DC system in terms of the conversion between Available Potential Energy (APE) and BPE represented by the APE dissipation rate, $\epsilon_p$. A positive value $\epsilon_p > 0$, means APE is dissipated (i.e., molecular diffusion smooths local density gradients converting APE to BPE). A negative value $\epsilon_p < 0$ means energy flows from BPE to APE. That is, molecular diffusion is not damping, but injecting energy into the APE reservoir. Tailleux \cite{tailleux_negative_2024} proposes that $\epsilon_p < 0$ is the signature of an active double-diffusive instability. This focus on the sign of the APE dissipation rate provides a formal energetic foundation for longstanding views of double-diffusive convection. That is, double diffusion gives rise to local inversions in the density gradient (as heat diffuses more rapidly than salt) to generate APE. For growing modes, this is converted to kinetic energy (KE) in convective overturning (a possible route to staircase formation). While Middleton and Taylor \cite{middleton_general_2020} focused on the integrated diapycnal buoyancy flux, Tailleux \cite{tailleux_negative_2024} argues the importance of the laminar term (internal energy exchange), associated with molecular energy transfer, in the APE dissipation rate, noting that the diapycnal buoyancy flux and internal energy exchange represent different components of a single conversion between APE and BPE. Also along these lines, Hieronymus and Carpenter \cite{hieronymus_energy_2016} demonstrate the importance of conductive transfer in the steady-state energy budget for diffusive convection.

Related to energy conversions, the total buoyancy flux (i.e., the sum of the negative temperature and positive salinity buoyancy fluxes) is negative, consistent with the buoyancy flux providing the source of KE. This corresponds to a net downward, or counter-gradient, density flux, or negative effective diffusivity ($K_{\rho} < 0$); double-diffusive convection strengthens the stratification.

While energy arguments help determine whether the system is in a state that permits instabilities to grow, they do not explain how such instabilities evolve into fully developed DC staircases. This process is examined in section \ref{mechanisms}.

\section{The DC interface and instability}
\label{interface}

\begin{figure}[h!]
\centering
\includegraphics[width=5in]{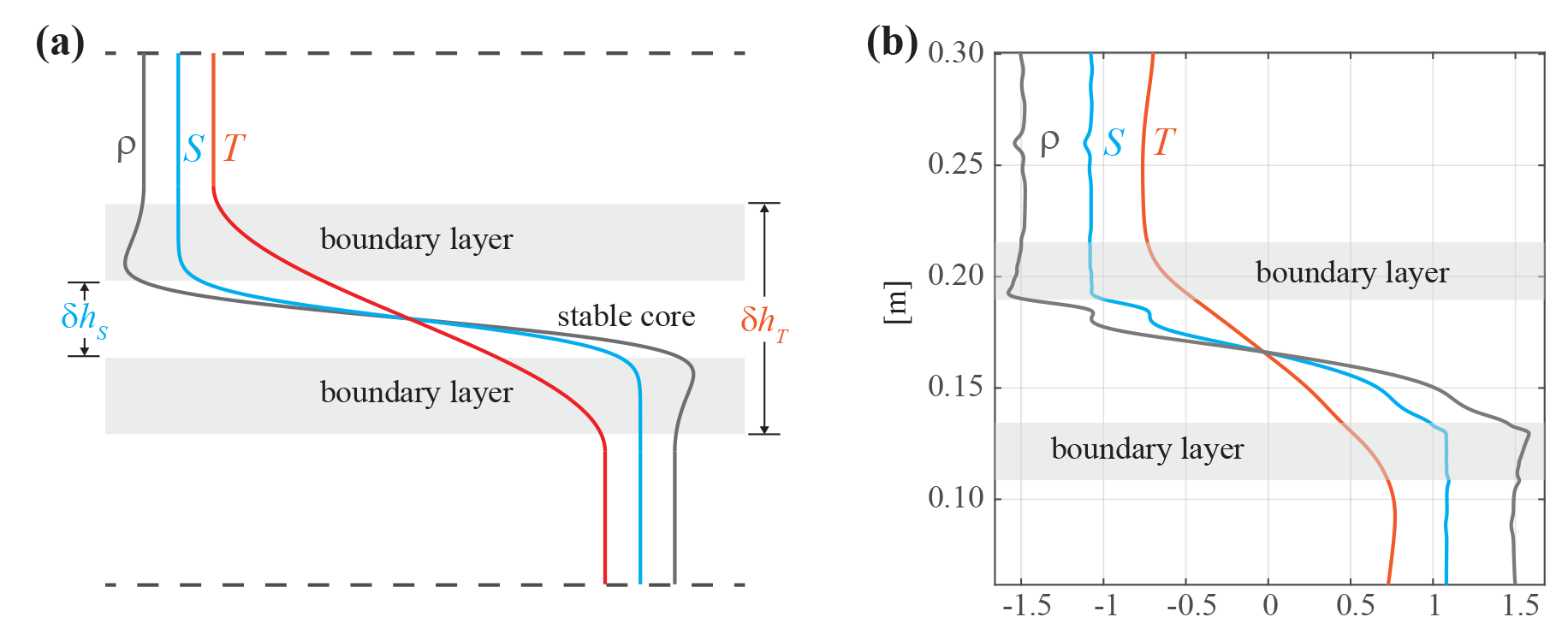}
\caption{Vertical structure of the DC interface. (a) Schematic profiles of temperature (red), salinity (blue), and density (grey) indicating the interface components: a gravitationally stable central core flanked by unstable diffusive boundary layers (shaded grey). The temperature interface thickness $\delta h_T$ is thicker than the salinity interface $\delta h_S$. (b) DNS profiles (arbitrarily scaled) showing the characteristic structure of a DC interface (simulation output from \cite{sommer_2014}). The simulation is characterized by interfacial density ratio $R_\rho = 6$, a thermal Rayleigh number (based on domain height) of $3.3 \times 10^6$, a Prandtl number of $Pr = 6.25$, and salt to heat molecular diffusivity ratio of $\tau = 0.01$.  The vertical domain height is 33~cm with periodic boundaries for perturbations from the background profiles.}
\label{interfacefig}
\end{figure}

Closely linked to the system energetics, a fundamental aspect of DC staircases is the structure of the interface that separates adjacent mixed layers. A DC interface is characterized by two gravitationally unstable boundary layers on either side of a stable core (Figure \ref{interfacefig}) \cite{linden_diffusive_1978}. The small diffusivity ratio $\tau = \kappa_S/\kappa_T \approx 0.01$ leads to a destabilizing thermal interface (thickness $\delta h_T$) that grows more rapidly than a salinity interface (thickness $\delta h_S$). This thickness difference ($\delta h_T > \delta h_S$) produces the unstable boundary layers (Figure \ref{interfacefig}), which are responsible for convective release.

The classical model of the DC interface is based on the balance between molecular diffusion in the interface and convective transport that arises when the unstable fluid is released from the boundary layers on either side of the core and swept into the mixed layers, after which the boundary layers begin to grow again. Assuming this balance in a one-dimensional analytical model, Linden and Shirtcliffe \cite{linden_diffusive_1978} showed that a steady-state DC interface can only be maintained if $R_{\rho} < \tau^{-1/2}$, where $R_{\rho}$ is an interfacial density ratio based on mean $T$ and $S$ gradients across the interface core. This criterion specifies the conditions for a {\it{steady-state}} balance between diffusive and convective fluxes. With oceanic values of $\tau \approx 0.01$, the criterion suggests that steady-state DC interfaces will not persist when the system has $R_{\rho} \gtrsim 10$. 

In the regime $R_\rho \gtrsim 2$, direct numerical simulations (DNS) confirm the existence of the double-boundary-layer structure of a DC interface, including the presence of the stable core across which fluxes of $T$ and $S$ are molecular \cite{carpenter_simulations_2012}.  Although mixed layer turbulence and circulation cells often disturb the boundary layers, simulations and observations have also demonstrated that $\delta h_T$ is statistically larger than $\delta h_S$, with a relatively undisturbed core for $R_\rho \gtrsim 2$ \cite{sommer_2014}. Simulations show how the unstable boundary layer fluid is swept from the stable core through convective cells in the mixed layers. However, no universal boundary-layer Rayleigh number threshold governing boundary-layer growth has been identified. For $R_{\rho} > \tau^{-1/2}$, the interface can persist, but not in steady-state; the stabilizing salinity gradient is too strong for mixed-layer convection to continuously compensate for diffusive thickening of the interface core. Worster \cite{worster_time-dependent_2004} bridges high and low $R_{\rho}$ regimes using a time-dependent model that does not assume marginal boundary-layer stability, rather a variable boundary-layer Rayleigh number reflects the competition between diffusive growth and episodic convective release.

\section{DC staircase formation mechanisms}
\label{mechanisms}

Several mechanisms have been proposed to account for DC staircase formation. We review them in this section and highlight their commonalities and differences. While many of the results reviewed here have been derived in effectively one-dimensional configurations, their application to the ocean is justified in many settings by the observed geometry of oceanic staircases. As discussed in section 6, oceanic staircases are frequently characterized by large lateral aspect ratios; layer properties vary much more rapidly in the vertical than in the horizontal, and vertical flux divergences are expected to dominate the structural evolution of the staircase. Although exceptions may exist (e.g., an example below of a staircase with high lateral variability), a one-dimensional framework generally provides an appropriate basis for understanding staircase maintenance and layer merging, even in the presence of lateral variability.

\subsection{Applied flux mechanism}
\label{appliedflux}

Laboratory experiments in which a stable salinity gradient is heated from below informed the original understanding of DC staircases \cite{turner_new_1964}. This setup generates a series of horizontal convecting layers, through the so-called applied flux mechanism of DC staircase formation. The first layer forms at the bottom, and a new layer forms above the bottom layer when a Rayleigh number characterizing the thermal boundary layer above the convecting region exceeds a critical value. The applied flux mechanism offers insight into DC interface structure and Rayleigh number bounds, but fluxes in most ocean staircases are determined by the staircase structure itself rather than being externally imposed. Next, we review a mechanism for DC staircase formation that operates without an externally applied heat flux.

\subsection{Thermohaline-shear instability}
\label{thermohalineshear}

Linear stability analysis on constant DC background $T$ and $S$ gradients predicts instability for $R_{\rho} < (Pr+1)/(Pr + \tau)$, where $Pr = \nu/\kappa_T$ is the Prandtl number and $\nu$ is kinematic viscosity \cite{veronis1965finite}. For the ocean characterized by $Pr \sim 7$ and $\tau \sim$10$^{-2}$, this yields only the narrow range $R_{\rho} < 1.14$. The linear stability analysis results in an oscillatory unstable mode where a displaced fluid parcel more rapidly adjusts its temperature anomaly than its salinity anomaly, such that the resulting density anomaly is restoring and this drives an amplifying oscillation of the displaced parcel.
If the stabilizing salinity gradient is sufficiently strong, the small energy gained during each oscillation by the thermal transfer is unable to overcome viscous dissipation, and the instability is suppressed. If $R_{\rho} > 1.14$ viscous dissipation dominates the buoyancy-driven energy gain such that the growth of infinitesimally small perturbations is suppressed. However, as noted earlier, oceanic staircases are found where background $R_{\rho}$ can be several times larger than 1.14.

Radko \cite{radko_thermohaline_2016} addresses this discrepancy in an analysis that shows how the addition of weak vertical shear to the system (modeled as unidirectional shear flow with a sinusoidal velocity profile) 
can operate in conjunction with the DC instability to give rise to instability over a wider range of $R_{\rho}$. When rising and sinking parcels are advected horizontally by the shear, this can bring them into regions where vertical motions are enhanced. In this way, shear turns a slow, oscillatory process into a fast-growing instability. Linear stability analysis and DNS indicate shear flows can be unstable for systems with density ratios as high as $R_{\rho} = 10$ and Richardson numbers as large as 1000 \cite{radko_thermohaline_2016}. The energy source in this thermohaline-shear instability is close to pure DC in that most of it derives from the unstable temperature stratification. The nonlinear evolution to a DC staircase takes place when the growing perturbations become sufficiently large that they become density inversions and convective overturns creating mixed layers. Initially, thin layers are formed of thickness consistent with the wavelength of the shear \cite{radko_thermohaline_2016}. As the staircase evolves, the mixed layers continue to merge to an equilibrium state; we discuss the process in section \ref{evolution}.

For this thermohaline-shear instability, the background shear is steady and has a specific vertical structure. This framework has recently been extended to subcritical oscillatory shear, which can, via time-dependent shear instability, generate layers that may be precursors for a fully developed staircase \cite{radko_concealed_2026}. A more generic setting might be characterized by weak turbulence, considered next.

\subsection{Turbulence-driven formation}
\label{turbulence}

Ma and Peltier \cite{ma_thermohaline-turbulence_2022} present a mechanism for staircase formation, which is an extension of the Phillips mechanism \cite{phillips_turbulence_1972} for layer formation in a single-component system, that requires both DC stratification and weak background turbulence. The strength of background turbulence is characterized by the buoyancy Reynolds number $Re_b = \epsilon/(\nu N^2)$, where $\epsilon$ is the dissipation rate of kinetic energy, and $N^2 = -(g/\rho_0)\partial \rho/\partial z$, where $g$ is acceleration due to gravity and $\rho_0$ is a mean background density. The mechanism draws on the observation-based model of Bouffard and Boegman \cite{bouffard_diapycnal_2013} for turbulent diffusivities ($K_T$ and $K_S$) that differ for $T$ and $S$, and with functional forms that depend on the $Re_b$ regime. In Ma and Peltier's \cite{ma_thermohaline-turbulence_2022} setup, mean-field equations governing the large-scale $T$ and $S$ depend on $K_T$ and $K_S$, which depend on $Re_b$, which in turn depends on $N^2$ that changes with the evolution of $T$ and $S$. Their linear stability analysis reveals that the system is unstable, via a turbulent feedback mechanism that does not depend significantly on $R_{\rho}$, for $Re_b \lesssim 100$ \cite{ma_thermohaline_2022}.

The basic mechanism for the linear instability when $Re_b \lesssim 100$ is as follows. A perturbation to linear DC $T$ and $S$ profiles can give rise to flux divergences that amplify the resulting density perturbations. The turbulent diffusivity of salt $K_S$ increases more strongly than $K_T$ with increases in $Re_b$ in this regime \cite{bouffard_diapycnal_2013,ma_thermohaline_2022}. This results in a larger salinity flux divergence compared to the temperature flux divergence that leads to growth of the density perturbations. Analogous to the case of pure double-diffusion, the net turbulent diffusivity is negative ($K_{\rho} < 0$), and the individual diffusivities for temperature and salinity ($K_T$ and $K_S$) are both positive, reflecting the fact that each constituent diffuses down its own gradient. Middleton et al. \cite{middleton_reconciling_2025} unified multiple layer-generation mechanisms, including the Phillips \cite{phillips_turbulence_1972} and Ma and Peltier \cite{ma_thermohaline-turbulence_2022} mechanisms, into a general framework applicable to both single- and two-component stratified fluids, highlighting the role of negative turbulent diffusivity for buoyancy.

Because the thermohaline-turbulence instability theory is based on the mean-field equations, it cannot account for the very small scale behaviour, and the predicted growth rate increases without bound as the vertical wavenumber increases (i.e., the ultraviolet catastrophe). This issue arises because in the parameterization approach of this model, there is a separation between the layer scales and the smallest turbulence scales. A multiscale analysis (as done by Radko \cite{radko_thermohaline_2019} for the SF regime) is therefore required to regularize the ultraviolet catastrophe by extending the mean-field framework to include higher-order (hyperdiffusive) terms. This approach would explicate the physical mechanisms that stabilize the short wavelength limit, and yield physically consistent maximal growth rates and dominant layer thicknesses.

The applied flux, thermohaline-shear and turbulence-driven formation mechanisms discussed so far do not require lateral $T$ and $S$ gradients. Next, we describe a scenario for DC staircase formation that relies on lateral gradients. 

\subsection{Thermohaline intrusions}
\label{thermohalineintrusions}

Thermohaline intrusions are lateral interleaving flows that exhibit alternating DC and SF regions in depth and involve both vertical and lateral gradients in temperature and salinity. Their nearly lateral buoyancy-driven propagation is forced in part by vertical divergences/convergences of vertical double-diffusive buoyancy fluxes \cite{ruddick_oceanic_2003}. The frequent co-occurrence of thermohaline intrusions and DC staircases in the ocean suggests a relationship between the two.

For the SF regime, Merryfield \cite{merryfield_origin_2000} used a parametric flux model (that allows for coupling between cm-scale double-diffusive and km-scale lateral intrusive processes) to show that interleaving perturbations can evolve into either thermohaline intrusions or staircases. Extending this to the DC regime, Bebieva and Timmermans \cite{bebieva_relationship_2017} show that a DC staircase is predicted from interleaving perturbations if the vertical background density ratio is below a critical value ($R_{\rho} < R_{\rho}^{\text{cr}}$) while persistent intrusions are expected otherwise. The critical density ratio $R_{\rho}^{\text{cr}}$ depends on the vertical and lateral background $T$ and $S$ gradients as well as parameters of the parametric flux model including eddy diffusivity and viscosity. Physically, the different end states reflect how buoyancy anomalies generated by interleaving perturbations are relaxed. If $R_{\rho} < R_{\rho}^{\text{cr}}$, these anomalies are efficiently removed by vertical rearrangement, favoring staircase development. If $R_{\rho} > R_{\rho}^{\text{cr}}$, vertical adjustment is suppressed and the system preferentially exports buoyancy anomalies horizontally, resulting in intrusions.

In a related study, Bebieva and Timmermans \cite{bebieva_doublediffusive_2019} explored the conceptual link between intrusions and staircases with the hypothesis that a staircase represents the run-down state of laterally-propagating intrusions. 
The run-down would be associated with a counter-gradient density flux ($K_{\rho} < 0$) that increases the overall stratification, as also suggested by Garrett \cite{Garrett1982} in a simplified model of the diapycnal fluxes associated with stirring by mesoscale eddies.
The distance and speed of this run-down process are primarily controlled by the vertical stratification (stronger $N$ results in a shorter run-down distance), and the strength of the lateral buoyancy gradient (weaker lateral temperature gradients result in shorter run-down distances) \cite{bebieva_doublediffusive_2019}. 
To summarize, intrusions and staircases may be alternate end states of interleaving perturbations \cite{bebieva_relationship_2017}, or staircases may be the outcome of decaying intrusions \cite{bebieva_doublediffusive_2019}.

\section{Evolution of DC staircases}
\label{evolution}
Having put forward several plausible mechanisms for initial staircase formation, we now turn to consideration of how DC staircases evolve in the ocean setting after formation.

\subsection{Layer merging, splitting and equilibrium thickness}
\label{mergesplit}

After staircase formation, individual layers have been observed (e.g., in laboratory experiments \cite{huppert_heating_1979}) to evolve by growing in thickness via layer merging. Two possible merging patterns were  documented by Linden \cite{linden_formation_1976} and later formally classified by Radko \cite{radko_mechanics_2007} as a Buoyancy merger (B-merger) and a Height merger (H-merger). A B-merger takes place when an interface strengthens at the expense of an adjacent weaker interface causing two mixed layers to become one, while an H-merger takes place when an interface migrates vertically until it coalesces with another. B-mergers are found to be more common in the ocean \cite{radko_double-diffusive_2014}. In quiescent (low $Re_b$) environments characterized by relatively weak lateral gradients, layer merging (driven by vertical flux imbalances, the B-merger) is predicted from the stability analysis of a staircase with mixed layers of uniform thickness \cite{radko_mechanics_2007}. A requirement for instability (spontaneous merging of layers) is that the ratio of salt to heat fluxes (termed $\gamma$) decreases with increasing interfacial $R_{\rho}$ ($d\gamma/dR_{\rho} < 0$). Consider, for example, any initial interface perturbation such that an interface with larger $R_{\rho}$ is subject to a lower $\gamma$ than a weaker (smaller $R_{\rho}$) adjacent interface. In this case, the flux of the stabilizing salinity relative to the destabilizing flux of heat is reduced across the stronger interface, which strengthens the interface further such that it can support larger temperature and salinity differences. Conversely,  the weaker interface is subject to higher $\gamma$ (higher stabilizing salt flux relative to the heat flux), which enhances its erosion and leads to the collapse of the interface and merger of the adjacent mixed layers.

Ultimately, merging would stop at some maximal layer thickness where the stability criterion ($d\gamma/dR_{\rho} < 0$) no longer holds and a staircase would thereafter be considered to be in a stable state. While the concept is established theoretically, we do not have clear proof that a DC staircase (in the laboratory or the field) evolves to and remains in such a terminally stable state. In the range of $R_{\rho}$ characterizing observed DC staircases, $\gamma$ is often found to be constant (e.g., laboratory experiments indicate $\gamma \approx 0.15$ for $R_{\rho} > 2$ \cite{turner_coupled_1965}) or  increasing with $R_{\rho}$ \cite{carpenter_simulations_2012}, or the functional dependence of $\gamma$ on $R_{\rho}$ may be time dependent \cite{worster_time-dependent_2004}.
DNS suggest layer merging is important, with merging slowing over time \cite{noguchi_multi-layered_2010}. However, most simulations are limited by finite domain size, relatively low Rayleigh numbers, and often larger diffusivity ratios than the oceans.
Consistent with this lack of clear evidence for a unique terminal layer thickness, DNS by Yang et al. \cite{yang_layering_2022} demonstrate that, even for identical system parameters, sheared DC systems can evolve toward multiple distinct staircase configurations with different layer thicknesses. 

In contrast to an equilibrium layer thickness being set by merging cessation, Kelley \cite{kelley_explaining_nodate} suggests that this final layer thickness is set by a balance between layer merging and interface splitting. A Richardson number $Ri_{conv}$ based on the convective shear across an interface falls below a critical value when convection becomes sufficiently strong following mergers. Kelvin-Helmholtz instability ensues and a mixed region of fluid develops within the original interface creating new thinner layers (termed {\it{sandwich}} layers). This interface splitting process increases $Ri_{conv}$ above the critical value. A stabilization of staircase layer thickness is suggested to arise because the dynamic competition between layer merging and interface splitting ensures $Ri_{conv}$ is confined to marginally stable values. To the best of our knowledge, interface splitting has not been observed in DNS. We discuss observational context for this process in section \ref{ocean}.

\subsection{The influence of background turbulence}
\label{turbulence}

Background turbulence can play a dual role in staircase dynamics. In the ocean interior, this turbulence is predominantly driven by internal wave breaking, particularly from near-inertial wave packets that intermittently increase vertical shear as they propagate \cite{Garrett1979}. While a modest level of turbulence may be essential for DC staircase formation, sufficiently strong turbulence can inhibit staircase development or disrupt existing staircases. Using DNS, Brown and Radko \cite{brown_disruption_2022} showed that staircases are disrupted when the Richardson number based on the background shear and stratification falls below a critical value, with thinner staircase mixed layers being more vulnerable than thicker ones. Shibley and Timmermans \cite{shibley_formation_2019} used a 1D model with parameterized turbulence to deduce a critical percentage of time that intermittent turbulence can persist beyond which DC interfaces cannot be maintained. These studies highlight that DC staircases can persist only within a window of turbulence intensity.

Ma and Peltier \cite{ma_diffusive-convection_2024} identified two regimes for a quasi-steady DC staircase ($R_{\rho}\lesssim\tau^{-1/2}$) depending on turbulence levels: a double-diffusion regime (when $Re_b \lesssim 10$), where the mixing is more accurately described as a differential diffusion process. In this regime, the molecular properties of $T$ and $S$ still have an influence because the turbulence is not sufficiently strong to completely overwhelm molecular effects. For this regime, the potential energy released from the $T$ field (negative buoyancy flux) balances viscous dissipation. There is also a hybrid regime (when $10 \lesssim Re_b \lesssim 100$) driven by both external turbulence and negative buoyancy fluxes and the staircase layers can be unstably stratified (instead of homogeneous), evidently the result of active scouring. In the hybrid regime, the turbulence is only strong enough to drive the required fluxes, but not sufficiently strong to compromise the  integrity of the DC interface with its unstable boundary layers. In sum, the persistence of a DC staircase depends on whether turbulence can enhance fluxes without destroying the DC interfaces that define the staircase.

\section{Oceanographic examples: Insights into staircase formation and variability}
\label{ocean}

In this section, we illustrate how the mechanisms described in previous sections manifest in the ocean. Rather than presenting an exhaustive survey of oceanic DC staircases, we focus on a set of environments that exemplify distinct pathways for DC staircase formation and evolution. 

\subsection{DC staircases in bottom waters: Heating from below}

\begin{figure}[h!]
\centering
\includegraphics[width=4.5in]{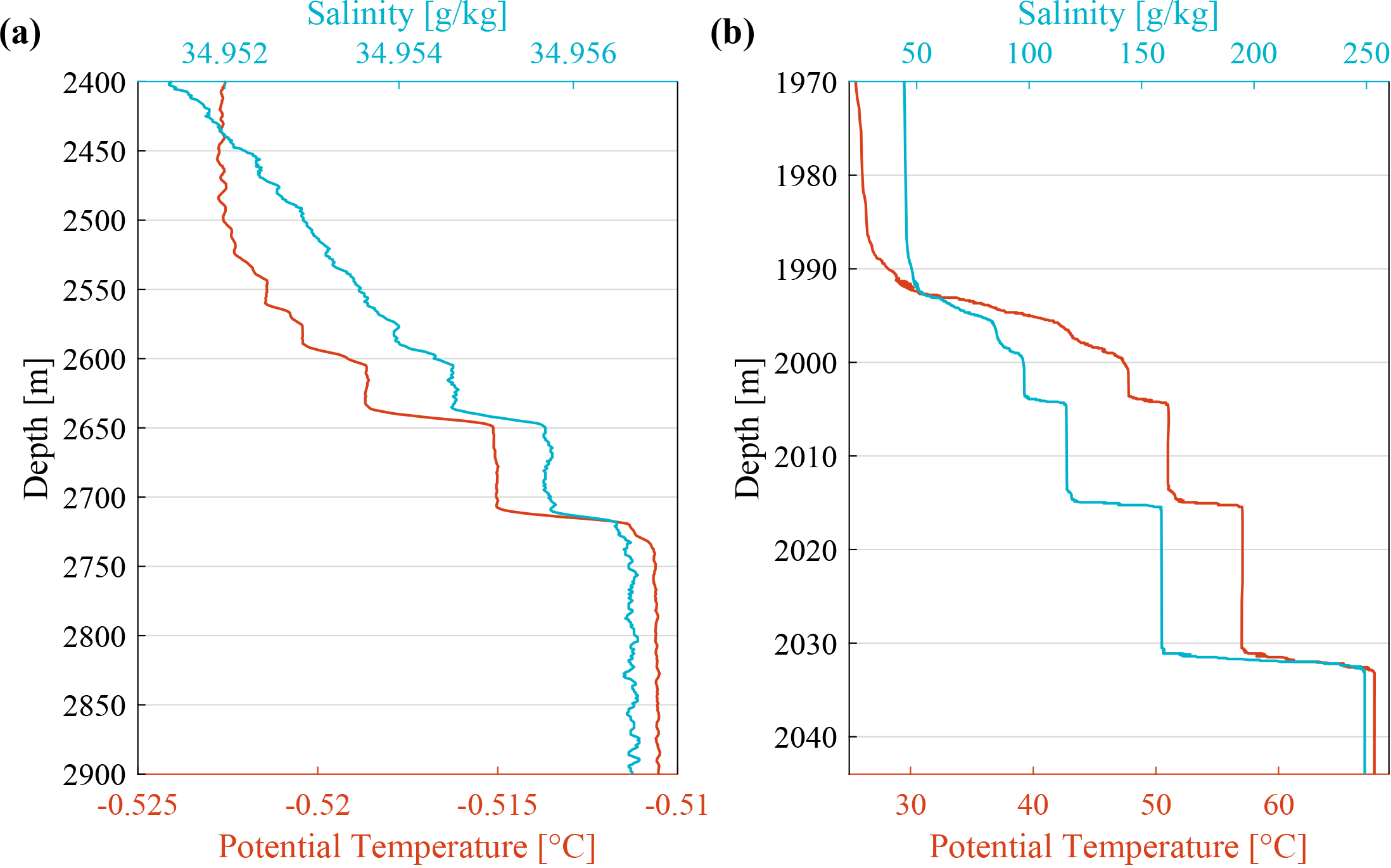}
\caption{Potential temperature and salinity profiles from (a) the Arctic Ocean's deep Canada Basin  and (b) the Red Sea's Atlantis II Deep \cite{swift_vertical_2012}. Arctic profiles were taken in August 2008 during the Joint Ocean Ice Study/Beaufort Gyre Observing System expedition \cite{arroyo_structure_2025} and Red Sea profiles were taken in October 2008 during R/V Oceanus Cruise 449-6 using a custom high-range sensor developed at Woods Hole Oceanographic Institution \cite{swift_vertical_2012}.}
\label{geothermalfigure}
\end{figure}

In the ocean's bottom waters, geothermal heat provides a persistent upward heat source and natural evidence for the staircase formation mechanism described in section~\ref{mechanisms}\ref{appliedflux}, in which a stable salinity gradient is heated from below. Two examples of bottom-heated DC staircases are found in markedly different environments, the Arctic Ocean and the Red Sea (Figure \ref{geothermalfigure}). In the isolated bottom waters of the Arctic Ocean's Canada Basin, with temperatures near freezing, a persistent staircase has been observed for decades \cite{timmermans_thermohaline_2003}. The staircase consists of several convective layers with thicknesses ranging from about 10-60~m, separated by 2-16~m thick interfaces (Figure \ref{geothermalfigure}a). Temperature jumps across interfaces are only a few millidegrees and salinity jumps are less than 1 part per million. The mean density ratio is small ($R_{\rho} \approx 1.6$). Timmermans et al. \cite{timmermans_thermohaline_2003} showed that the observed interfaces are too thick to support fluxes inferred from DC flux parameterizations. Such parameterizations are typically based on 4/3-flux laws \cite{radko_double-diffusive_2013}, which assume a non-rotating, low-turbulence environment. Carpenter and Timmermans \cite{carpenter_does_2014} subsequently speculated that these thick interfaces likely exist in a regime where the influence of planetary rotation significantly reduces heat fluxes below standard laboratory-derived predictions.

More than 6000~km south of the Canada Basin lies the Red Sea, a subtropical basin in which a DC staircase has also been observed to persist for decades in the bottom waters \cite{swift_vertical_2012}. Observations of Swift et al. \cite{swift_vertical_2012} reveal several convective layers, 5-120~m thick, separated by 1-8~m thick interfaces (Figure \ref{geothermalfigure}b). In contrast to the Arctic, $T$ and $S$ jumps across interfaces are extreme, ranging from a few to $\sim$10~$^\circ$C and 10-100 parts per thousand. The mean density ratio is $R_{\rho} \approx 10$. The heat source is hot, hypersaline brine that is believed to be discharged from hydrothermal vents to form a laterally uniform bottom layer with temperatures about 68$^\circ$C (Figure \ref{geothermalfigure}b). 

The Arctic and Red Sea systems illustrate the broad range of conditions under which DC staircases can form and persist, spanning orders of magnitude in temperature and salinity contrasts and density ratio. It remains unclear what controls layer thicknesses across such disparate regimes which have layer thicknesses of the same order, or whether these long-lived staircases are in equilibrium. Their time dependence, including adjustment timescale following disruptions, is unknown. Both systems present significant observational challenges owing to their significant depths, interface jumps that are barely within resolution limits (in the Arctic case) and extreme $T$ and $S$ ranges (in the Red Sea case). It is further unclear what sets the interface thicknesses, what is the role of turbulence and whether planetary rotation is important. Ultimately, it remains to be determined whether there is a unified dynamical framework for these two DC staircases, or whether they are distinct regimes that simply look similar.

\subsection{DC staircases and thermohaline intrusions}

\begin{figure}[!h]
\centering
\includegraphics[width=5.6in]{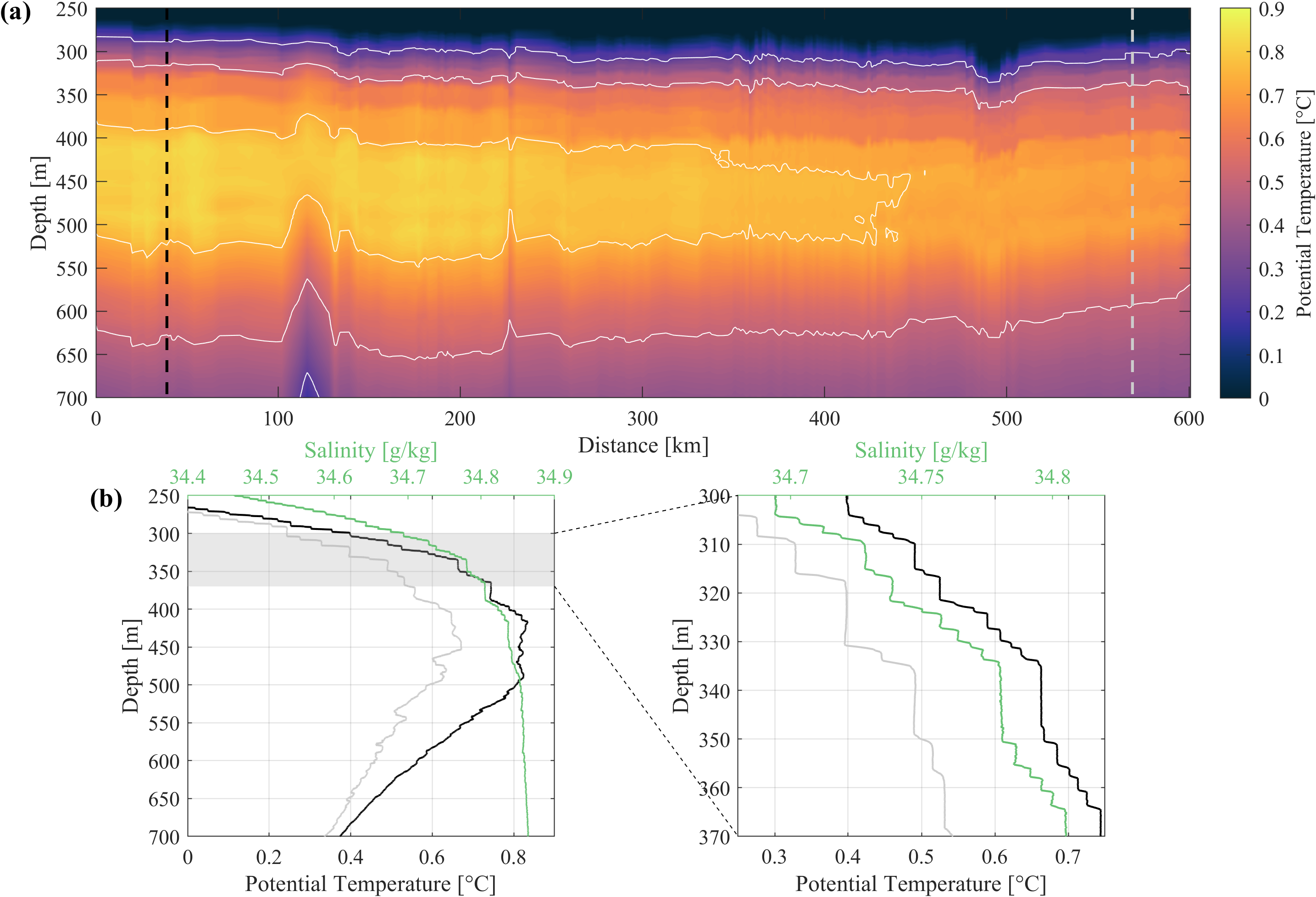}
\caption{Data from an Ice-Tethered Profiler (ITP) that drifted in the Arctic Ocean's Canada Basin in 2015 (ITP87). (a) Potential temperature ($^\circ$C) versus depth and distance along the ITP drift track from the northwestern boundary (left) to the southeastern basin interior (right) over the course of $\sim$one year. 
(b) Potential temperature and salinity vs. depth profiles, from the location of the black dashed vertical line in (a), illustrating the detailed DC staircase structure. The additional potential temperature profile shown in grey is at the location of the grey dashed vertical line in (a).}
\label{intrusions}
\end{figure}

\begin{figure}[h!]
\centering
\includegraphics[width=2.5in]{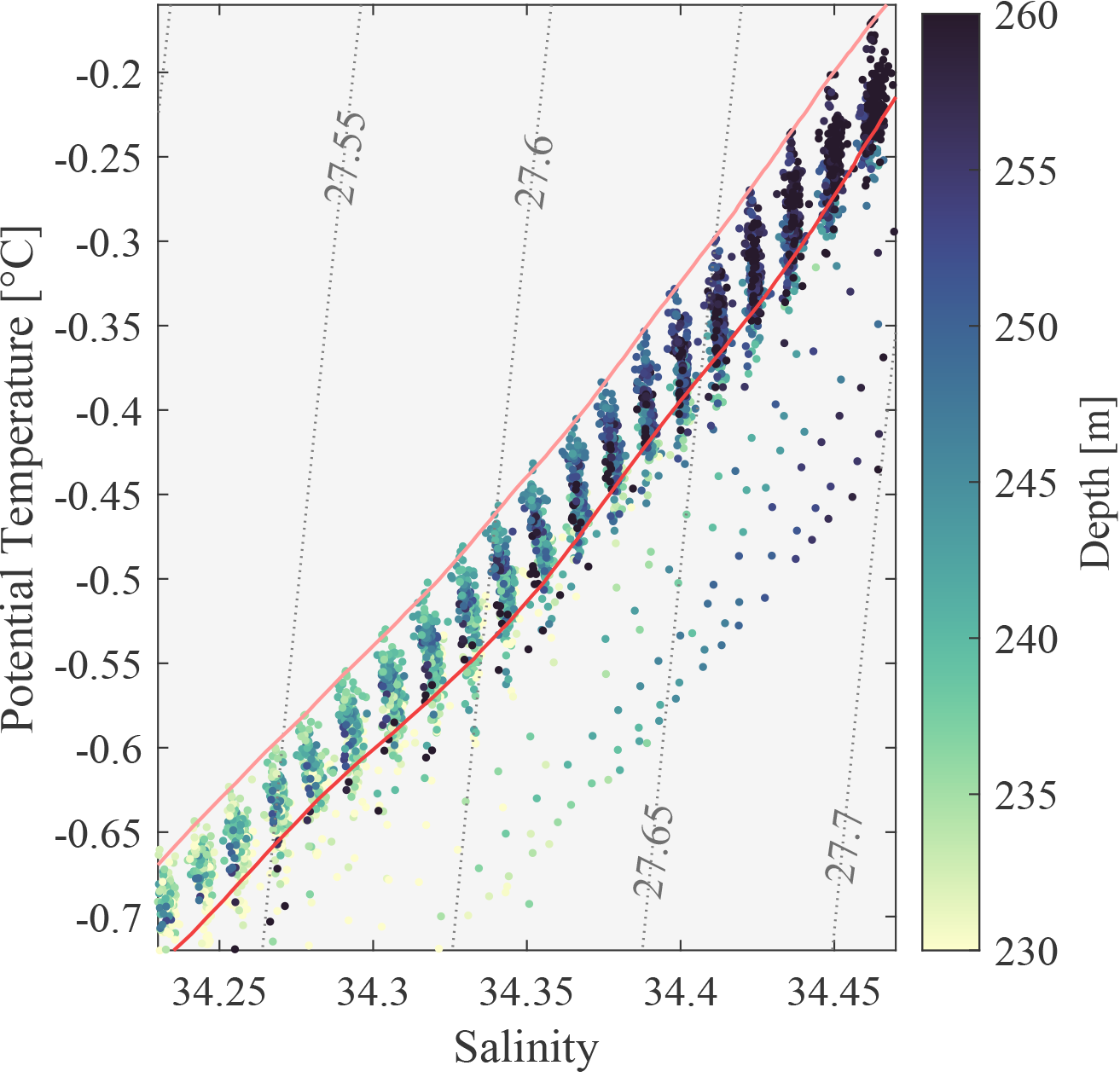}
\caption{Ice-Tethered Profiler (ITP 87) data showing potential temperature vs. salinity of the mixed layers in the DC staircase at the top boundary of the Atlantic Water layer in the Arctic Ocean. Colors indicate water-column depth, and isopycnals ($\rho - 1000$, kg~m$^{-3}$) are shown as black contours. The red lines show bounding profiles, with the light red profile located about 400~km to the southeast of the dark red profile.}
\label{TS}
\end{figure}

There is evidence in the Arctic Ocean for DC staircase formation associated with thermohaline intrusions (section \ref{mechanisms}\ref{thermohalineintrusions}). Relatively warm Atlantic Water propagates into the Canada Basin interior from its boundaries via thermohaline intrusions \cite{carmack_changes_1997} centered  $\sim$450~m depth (Figure \ref{intrusions}). The Atlantic Water cools from the northwestern boundary through the southeastern basin interior, with heat propagating laterally in these intrusions (Figure \ref{intrusions}a). Overlying the intrusions is a DC staircase (Figure \ref{intrusions}b), with $O(1)$~m–thick layers that extend laterally for hundreds of kilometers, implying aspect ratios of $O(10^6)$ \cite{timmermans_icetethered_2008}. In $T$–$S$ space, properties of individual mixed layers form clusters with lateral $T$ and $S$ gradients that are consistent with a steady-state balance between advection along layers and vertical divergences of DC heat and salt fluxes \cite{timmermans_icetethered_2008} (Figure \ref{TS}). 


As discussed in section \ref{mechanisms}\ref{thermohalineintrusions}, linear theory predicts that interleaving perturbations may evolve to either a DC staircase or thermohaline intrusions depending on vertical and lateral property gradients. While Canada Basin observations have been shown to be broadly consistent with theory, the theory cannot be used for prediction because of the strong dependence on poorly-constrained eddy diffusivity \cite{bebieva_relationship_2017}. Further, linear theory cannot account for the observed finestructure, including thin mixed layers embedded between thicker staircase layers (Figure~\ref{intrusions}), which may influence interpretations of staircase/intrusion relationships.

DC staircases have alternatively been proposed as the run-down state of thermohaline intrusions \cite{bebieva_doublediffusive_2019}, which introduces a potential inconsistency. The model predicts that Canada Basin intrusions should only propagate a few kilometers into the basin before transitioning to a DC staircase \cite{bebieva_doublediffusive_2019}. However, the staircase is observed to extend for hundreds of kilometers across the basin, with strong coherence of individual layers (Figures \ref{intrusions} and \ref{TS}). If DC staircases are the result of run-down intrusions, it is unclear why they are observed so far from basin boundaries where intrusions originate. This lends support for staircase formation from perturbations to the background stratification \cite{bebieva_relationship_2017}, although as noted this interpretation has its own limitations.

\subsection{DC staircase evolution: evidence for merging and splitting}

\begin{figure}[h!]
\centering
\includegraphics[width=5.5in]{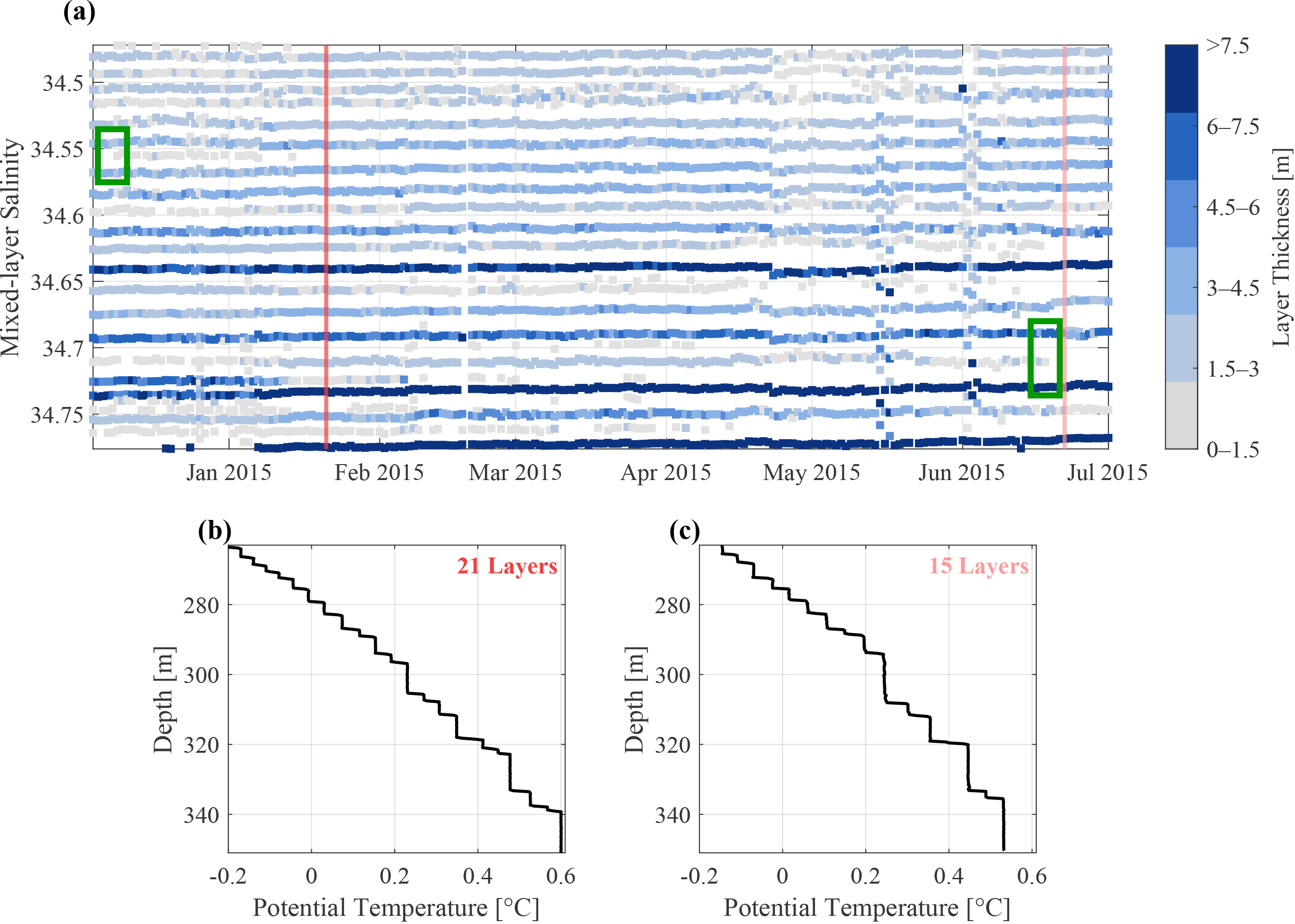}
\caption{Ice-Tethered Profiler data showing the Canada Basin DC staircase. (a) Salinity of mixed layers in a portion of the staircase over time, where each detected mixed layer in a vertical profile is shown as a colored square, with color indicating layer thickness. During the period shown (4 Dec 2014 to 1 Jul 2015), the ITP drifted about 600 km from the northwest to southeast Canada Basin. Consequently, it is not possible to unambiguously separate temporal evolution (e.g., active merging or splitting) from spatial variability. The green boxes indicate representative examples where horizontal layer clusters are seen to emerge (potential splitting, left box) or disappear (potential merging, right box).
Red vertical lines mark the two example profiles highlighted in (b) and (c). (b–c) Potential temperature versus depth for the left (b) and right (c) example profiles, with the total number of layers annotated in the top-right of each subplot. The decrease from 21 to 15 layers between these profiles is consistent with layer merging. Mixed layers were identified in each profile as continuous depth intervals where the vertical gradient of potential temperature is small, using k-means clustering to delineate low-gradient regions. Note that the profiles shown here are from high-resolution ITP profiles in a quiescent environment. In other settings, layers may fail to be detected even when present, for example, when their vertical scale is smaller than the vertical sampling resolution of the platform \cite{van_der_boog_double-diffusive_2021}, or when turbulence reduces the sharpness of interfaces \cite{radko_concealed_2026}.}
\label{layertrack}
\end{figure}

In the ocean it is often unclear whether observed staircases are in equilibrium (e.g., set by a balance between layer merging and possibly interface splitting, section \ref{evolution}\ref{mergesplit}) or whether they are continually adjusting following intermittent disruptions \cite{brown_disruption_2022} or long-term shifting environmental conditions \cite{lundberg_climate_2025}. Direct ocean observations of merging and splitting are rare \cite{radko_double-diffusive_2014}, although this may be a reflection of observing limitations, including difficulty detecting sandwich layers (i.e., thin layers between thicker layers). High vertical-resolution observations nevertheless suggest that sandwich layers may be more common than previously recognized. Such features are apparent in Arctic DC staircases (e.g., Figure \ref{intrusions}b) and appear in time-depth records. For example, in Figure \ref{layertrack}a, near-horizontal clusters indicating mixed layers are seen to emerge or disappear over time. A comparison of two profiles separated by months and hundreds of kilometers (Figure \ref{layertrack}b,c) shows that the overall staircase structure is preserved, while the earlier profile contains 1.4 times as many layers (see also Figure \ref{layertrack}a), consistent with layer merging. 
Due to the Lagrangian nature of the Ice-Tethered Profiler drift, it is not possible to deduce unambiguously whether thin layers arise due to interface splitting, or whether the staircases are simply formed with multiple thin layers that then merge, leaving observed staircases at different stages of equilibration in a spatially and temporally varying field.

Determining conditions for merging/splitting will be essential for testing mechanisms for staircase evolution. Ocean merging examples have indicated consistency with the secondary merging instability \cite{radko_double-diffusive_2014}, although alternative processes cannot be ruled out. The formation of sandwich layers via the Kelvin–Helmholtz instability could provide an explanation for why ocean values of $Ri_{conv}$ are often found in the range of marginal stability ($\sim$0.1 to 1) \cite{kelley_explaining_nodate}. However, it may be difficult to distinguish interface splitting from lateral advection of different water masses \cite{kelley_explaining_nodate}. Consistent with this ambiguity, features resembling sandwich layers have been observed in DNS although they were found to result from lateral convergences of interfacial fluid \cite{sommer_2014}. 

\subsection{DC staircases in turbulent environments}

DC staircases can persist in the presence of weak or intermittent turbulence, but are disrupted when turbulence becomes sufficiently strong (section \ref{evolution}\ref{turbulence}). For example, although the Arctic Ocean is often viewed as energetically quiescent, which allows for the persistence of the DC staircase across hundreds of kilometers, some level of turbulence may nevertheless be necessary for its formation. Ma and Peltier \cite{ma_thermohaline-turbulence_2022} perform DNS in a configuration relevant to the Arctic DC staircase which confirm instability in the $Re_b \lesssim 100$ range, supporting the existence of a hybrid regime in which double diffusion and turbulence coexist. For the same setting, Shibley and Timmermans \cite{shibley_formation_2019} infer that a persistent staircase will only be observed when shear-driven turbulence is active less than 15\% of the time. DC staircases have been shown to break down at basin boundaries where turbulence is enhanced (e.g., the deep Canada Basin staircase (Figure \ref{geothermalfigure}) loses its structure at the basin margins \cite{timmermans_thermohaline_2003,arroyo_structure_2025}). 

Staircase disruption is also associated with energetic mesoscale eddies. In the example shown in Figure \ref{layertrack}a, the loss of lateral layer coherence around 1 June coincided with an anticyclonic eddy $\sim$15~km in diameter and centered $\sim$200~m depth. The deep portion of the eddy appears as isotherm displacements in Figure \ref{intrusions}a near 500~km (with isohalines, not shown, similarly displaced). Detailed observations of a different eddy embedded within the Arctic DC staircase show that the staircase is absent at the eddy flanks, where geostrophic shear is strongest and the geostrophic Richardson number drops below $O(100)$, indicating that even relatively weak shear can disrupt layering \cite{bebieva_examination_2016}.

\begin{figure}[!h]
\centering
\includegraphics[width=3.5in]{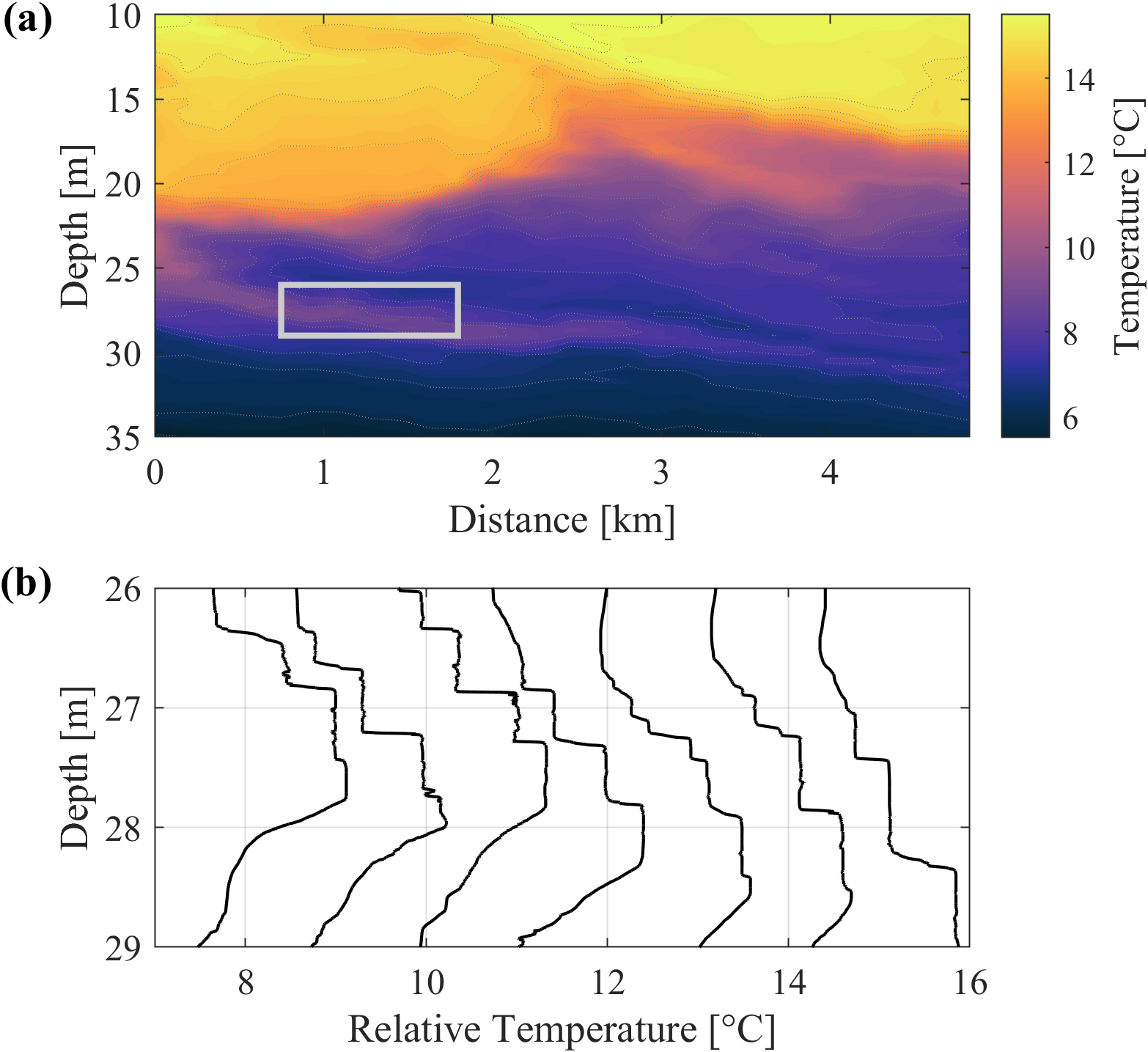}
\caption{Data from a glider transect in the Baltic Sea in June 2016 \cite{umlauf_diffusive_2018}. (a) Temperature ($^\circ$C) versus depth and distance along the transect measured by a microstructure sensor on the glider over a $\sim$10-hr period.
(b) Microstructure temperature profiles through a DC staircase (where each profile is successively offset by 1.2~$^\circ$C) in the region of the box shown in (a).}
\label{BalticGlider}
\end{figure}

Despite these disruptions, DC staircases can exist in dynamically active environments. In the Baltic Sea, for example, a DC staircase has been observed at the upper boundary of a warm, salty interleaving inflow only a few meters thick but extending laterally for several kilometers \cite{umlauf_diffusive_2018} (Figure \ref{BalticGlider}). Temperature jumps across staircase interfaces are $\sim$0.5$^\circ$C, two orders of magnitude larger than typical for the Arctic DC staircase. The staircase is localized and short-lived, persisting over scales of hundreds of meters and hours to days, rather than exhibiting the basin-scale coherence seen in the Arctic. This setting potentially corresponds to one where staircase formation is via the turbulence-driven or thermohaline-shear mechanisms. 
However, its formation and evolution are strongly time dependent, with no long-term equilibrium state. The Baltic Sea example highlights how DC staircases can occur even in rapidly evolving conditions.

\section{Summary and open questions}
\label{summary}

While linear theory provides valuable insight for most fluid instabilities, the development of a DC staircase is often found far from the predictions of linear stability theory. In addition to background DC stratification, other factors such as a sustained external heat source, turbulence within an optimal range, shear, and lateral $T$ and $S$ gradients may be required for DC staircase formation. After formation, DC staircases can be long-lasting coherent structures, maintained by sustained double diffusion and weak background turbulence, or short-lived transient features. They may evolve via layer merging and interface splitting or be disrupted by sufficiently strong turbulence. Perhaps the most fundamental scientific challenge in understanding DC staircases in dynamic oceanic settings lies in bridging the gap between the smallest-scale instabilities and the basin-scale persistent layers.

There are many open questions regarding the formation, equilibration and large-scale organization of DC staircases in the ocean. The relevance of different formation pathways remains uncertain. For example, the role of weak background turbulence in enabling staircase formation remains poorly constrained by quantitative observations. Further, the extent to which DC staircases represent the run-down state of thermohaline intrusions versus the outcome of perturbations to the background vertical and lateral stratification is not understood. If staircases arise from instabilities of local stratification, it remains unresolved how they achieve and maintain lateral coherence over basin scales. After a staircase forms, it is unclear whether equilibration is via marginal stability of sheared interfaces, through layer merging via secondary instability of an established staircase, or whether an equilibrium exists at all. The time-dependent behavior of staircases is not well constrained. For example, how rapidly does a staircase re-form after disruption (by a mesoscale eddy, for example), and over what timescales does re-equilibration occur? Closely related is the question of what controls layer thickness, and why comparable layer scales emerge across disparate oceanic regimes. Addressing these questions will require targeted ocean observations that can capture both finestructure and basin-scale variability, in conjunction with numerical simulations that can encompass a broader range of scales and parameter space. 

Finally, we close with a comment on the role of diffusive convection in global ocean mixing. While double diffusion can have an important influence regionally (such as in the Arctic Ocean), van der Boog et al. \cite{van_der_boog_double-diffusive_2021} estimate that double-diffusive mixing (both SF and DC) associated with staircases contributes $\lesssim$~1~\% of the energy required for global ocean mixing. However, double diffusion can be active even when staircases are not present, and has been proposed to account for significant levels of oceanic background turbulence \cite{middleton_2021}. In global ocean models, when local conditions have density ratios that meet the criteria for DC or SF, parameterized forms of $K_T$ and $K_S$ are incorporated. Ma and Peltier \cite{ma_diffusive-convection_2024} point out that a limitation of this approach is that turbulence and diffusive convection may be fundamentally related, and recommend an alternative approach that accounts for the interaction. Further work is needed to clarify the importance of differential mixing of heat and salt in the global oceans, both in the presence and absence of staircases, including to assess the cumulative influence of intermittent DC staircases (such as the Baltic Sea example presented here) that may occur frequently but remain under sampled.

\ack{We thank the Isaac Newton Institute for Mathematical Sciences, Cambridge, for support and hospitality during the programme Anti-diffusive dynamics: from sub-cellular to astrophysical scales. Support was also provided by the National Science Foundation Office of Polar Programs and the Office of Naval Research under the Multidisciplinary University Research Initiative. Support from the Helmholtz Association is also acknowledged.  \textbf{Data sources}: Deep Canada Basin data were collected by the Beaufort Gyre Exploration Program (\url{www2.whoi.edu/site/beaufortgyre/}) in collaboration with researchers from Fisheries and Oceans Canada at the Institute of Ocean Sciences. Ice-Tethered Profiler (ITP) data were provided by the ITP Program based at the Woods Hole Oceanographic Institution, WHOI (\url{www.whoi.edu/itp}, \cite{Krishfield2008ITP,Toole2011ITP}). The Red Sea data were provided by Amy Bower and Heather Furey (WHOI). Hydrographic data from the World Ocean Database (WOD) were also used \cite{Mishonov2024WOD}. We thank Chris Garrett, Leo Middleton, and one anonymous reviewer for careful and constructive comments that improved the manuscript.
}

\bibliography{doublediffusion.bib}

\end{document}